\documentclass{ismdproc}

\usepackage{cite,./mcite}

\begin{document}

%------------------------------------
\title{Phenomenology with unintegrated\\ parton showers\footnote{Preprint number: IFT-UAM/CSIC-11-21}}

% for the author list please adhere to the format of one of the following
% three examples

% use the following for a single author
%
%\author{{\slshape Joe Smith}\\[1ex]
%University of Antwerp, Groenenborgerlaan 171, 2020 Antwerpen, Belgium }

% use the following for several authors
%
%\author{{\slshape Jean Meunier$^1$, Ruth Miller$^2$,
%    Gerd M\"uller$^3$\footnote{Speaker}, Joe Smith$^3$}\\[1ex]
%$^1$CERN, 1211 Gen\`eve 23, Switzerland\\
%$^2$Fermilab, P.O. Box 500, Batavia, IL 60510-0500, USA\\
%$^3$DESY, Notketra{\ss}e 85, 22607 Hamburg, Germany}

% use the following for an author speaking on behalf of a collaboration
%
\author{{\slshape Michal De\'ak}\\[1ex]
Instituto de F\'isica T\'eorica UAM/CSIC, 
C/ Nicol‡s Cabrera 13-15,\\
Universidad Aut\'onoma de Madrid, 
Cantoblanco, Madrid 28049, SPAIN}

% please do not modify the following 5 lines
%\contribID{xy}  % will be entered by the editors
%\confID{yz}
%\acronym{ISMD2010}
%\doi            % will be entered by the editors

\maketitle

\begin{abstract}
We introduce a backward evolution Monte Carlo algorithm implementation of the CCFM equation and present latest developments in phenomenology of hadron-hadron collisions for the Monte Carlo generator {\sc Cascade}.
\end{abstract}

\section{Introduction}

The BFKL%~\cite{BFKL1,*BFKL2,*BFKL3,*BFKL4} 
and the CCFM
%~\cite{CCFM1,*CCFM2,*CCFM3} 
equations offer possibility to formulate so called unintegrated parton density functions (uPDFs). In uPDFs, in contrast to parton density functions (PDFs), the dependence on the transversal momentum of parton ${\bf k}$ is preserved. In PDFs, also sometimes called integrated parton density functions, the transversal momentum of the evolved parton density is integrated over, but on the other hand there is a dependence on a scale to which the PDF is evolved. The corresponding equation for evolution of PDFs is the DGLAP
%~\cite{DGLAP1,*DGLAP2,*DGLAP3,*DGLAP4} 
equation. Both uPDFs and PDFs depend on the longitudinal momentum fraction carried by the parton $x$. The latter gives us also a hint for relation between uPDFs - $\mathcal{F}$ and PDFs - $f$ which can be sketched by equation
\begin{equation}\label{eq:uPDFs}
xf(x,\mu)=\int\limits_{0}^{\mu}d^2{\bf k}\;\mathcal{F}(x,{\bf k})\;.
\end{equation}

Equation~\eqref{eq:uPDFs} shows that $\mathcal{F}(x,{\bf k})$ contains more information than the $f(x,\mu)$, one can therefore expect that uPDFs will be very interest for phenomenology. Note that the uPDFs obtained using the CCFM equation depend on the longitudinal momentum fraction $x$, the transversal momentum ${\bf k}$ and a scale $\mu$. Indeed, particular studies~\cite{Deak:2009xt,Deak:2008ky,Baranov:2007np} show that uPDFs, by looking on observables connected to transversal momentum of the final state particles, effectively contain information from higher orders of perturbation theory.

One of the most powerful methods, for solving evolution equations and obtain phenomenological results, is the probabilistic interpretation of their kernels and their implementation in Monte Carlo programs in a form of parton shower generators. In Monte Carlo programs every term in the expansion in $\alpha_S$ of the solution of the evolution equation is interpreted as a chain of parton emissions. The solution is obtained by summing all relevant terms and integrating them over free kinematic variables. In the previous paragraph we mentioned two different approaches to partonic content of the proton. From the point of view of a parton shower generator the difference is not so obvious. In a Monte Carlo program, which is solving the DGLAP equation and using PDFs as an initial condition, the transversal momentum of the last parton in the chain can be left free from integration and one can effectively obtain an uPDF which will also depend on the scale $\mu$ (situation is similar to the one for the CCFM uPDFs). Most of the differences between the approaches with PDFs and uPDFs is in the dynamics, small x resummation in the BFKL and the CCFM equations, but also in the way how is the transversal reflected in the kinematics of the parton. In the BFKL and the CCFM equation the partons are off-shell by $k^2=-{\bf k}^2$. In the DGLAP approach partons are kept on-shell which requires reshuffling of components of momenta~\cite{Nagy:2007ty}.

It seems that the truly unintegrated approach is more consistent and offers also richer dynamics. We will describe theoretical foundations and phenomenological results Monte Carlo generator {\sc Cascade}~\cite{CASCADE2} based on the CCFM equation.

\section{The CCFM equation for a Monte Carlo generator}

In the limit when the longitudinal proton momentum fraction $x$ carried by the parton is  very small, $x\ll 1$, the proton structure is dominated by the gluon component. The leading order in the BFKL equation includes only gluons. The CCFM equation can be formulated for the gluon uPDF, just by extending the BFKL equation by $1/(1-z)$ term in the splitting function and angular ordering of gluon emissions. There is also a formulation of the CCFM equation for the valence quark uPDF which we will return to later.

The probability density for a splitting of a gluon into two gluons with angular ordering constraint will be (without inclusion of small-$x$ virtual corrections)
\begin{equation}
d\mathcal{P}^{\theta}_i=\frac{\alpha_S}{2\pi}dz_i\frac{d^2{\bf q}_i^{\prime}}{{{\bf q}_i^{\prime}}^2}\hat{P}_{gg}(z_i)\Theta(|{\bf q}_i^{\prime}|>z_{i-1}|{\bf q}_{i-1}^{\prime}|)\Theta(1-z_i-\epsilon)
\end{equation}

with the first $\Theta$-function forcing angular ordering and the second introducing an infrared regulator $\epsilon$, which can be shown has to be $\epsilon=\frac{Q_0}{|{\bf q}^\prime|}$~\cite{Marchesini:1987cf}, with $Q_0$ being a constant infrared cut-off. Variable ${\bf q}_i^{\prime}={\bf q}_i/(1-z_i)$ is introduced as the transverse momentum of the emitted parton rescaled by factor $1/(1-z_i)$. By $\hat{P}_{gg}(z)$ we denote the gluon to gluon splitting function devoid of its non-singular terms.

Virtual corrections have to be included also into the splitting function used in the CCFM equation. The difference between the virtual correction factors of the BFKL and the DGLAP is that to be consistent one should apply the angular ordering condition also for the virtual corrections included in them.

The Sudakov form factor will read
\begin{equation}\label{eq:Sud}
\Delta_S({{\bf q}^{\prime}}^2_i,(z_{i-1}{\bf q}^{\prime}_{i-1})^2)=\exp\Bigg(-\int\limits_{(z_{i-1}{\bf q}_{i-1}^{\prime})^2}^{{{\bf q}_{i}^{\prime}}^2}\frac{d^2{\bf q}^{\prime}}{{{\bf q}^{\prime}}^2}\int\limits_{0}^{1-\frac{Q_0}{|{\bf q}^\prime|}}dz\frac{\alpha_S}{\pi}\frac{N_C}{1-z}\Bigg).
\end{equation}

The Non-Sudakov form factor is
\begin{equation}\label{eq:Non-Sud}
\Delta_{NS}({\bf k}^2_i,(z_{i-1}{\bf q}^{\prime}_{i-1})^2)=\exp\Bigg(-\int\limits_{(z_{i-1}{\bf q}_{i-1})^2}^{{\bf k}^2_i}\frac{d{\bf q}^2}{{\bf q}^2}\int\limits_{z_i}^{1}dz\frac{\alpha_S}{\pi}\frac{N_C}{z}\Bigg)
\end{equation}

in analogy to Regge form factor used in the BFKL equation respectively.

The CCFM equation reads
\begin{equation}\label{eq:IS-CCFM}
\begin{split}
\mathcal{F}(x,{\bf k},{{\bf q}^{\prime}}^2)&=\mathcal{F}(x,{\bf k},{{\bf q}_0^{\prime}}^2)+
\int\limits_{{{\bf q}_0^{\prime}}^2}^{{{\bf q}^{\prime}}^2}\frac{d^2\bar{\bf q}^{\prime}}{\bar{\bf q}^{\prime 2}}\frac{N_C\alpha_S}{\pi}\\
&\int\limits_{x}^{1-\frac{Q_0}{|{\bf q}^{\prime}|}}\frac{dz}{z}\mathcal{F}(x/z,{\bf k}^{\prime},\bar{\bf q}^{\prime 2})\Bigg(\frac{\Delta_{NS}({{\bf k}^{\prime}}^2,(z\bar{\bf q}^{\prime})^2)}{z}+\frac{1}{1-z}\Bigg)\Delta_S({{\bf q}^{\prime}}^2_{0},(z\bar{\bf q}^{\prime})^2),
\end{split}
\end{equation}
%\begin{equation}
%{\bf J}^2(q)=-{\bf t}_I({2\bf t}_1W_{1I}^I+{2\bf t}_2W_{2I}^I)-2{\bf t}_1({\bf t}_1W_{1I}^{1}-%{\bf t}_2W_{2I}^{2}-\frac{1}{2}{\bf t}_{2}W_{12}^{1})-2{\bf t}_2({\bf t}_1W_{1I}^{1}-{\bf %t}_2W_{2I}^{2}-\frac{1}{2}{\bf t}_{1}W_{12}^{1})
%\end{equation}

where ${\bf k}^\prime={\bf k}+{\bf q}$. The presence of the term responsible for soft gluon emissions in the CCFM splitting function provides summation  of logarithms of ${{\bf q}^\prime}^2/{{\bf q}^\prime}^2_0$ in addition to the $1/x$ logarithms summed by the BFKL equation in leading logarithmic precission.

The Sudakov form factor~\eqref{eq:Sud}, which gives the probability of no emission between to values of the evolution scale, is used to determine the value of the evolution scale for the next emission.

Appearance of the Non-Sudakov form factor~\eqref{eq:Non-Sud} and gluon virtuality brings technical difficulties in backward evolution formulation of the parton shower. A solution is presented in~\cite{CASCADE2}.

\section{Fits of uPDFs}

In~\cite{Bacchetta:2010hh} this parameterisation was used to fit initial CCFM gluon uPDF at the starting scale $Q_0=1.2~GeV$
\begin{equation}
x A_0(x,k_t)=N x^{-B} (1-x)^C (1-D x) e^{-(k_t-\mu)^2/\sigma^2}
\end{equation}
where $N,B, C, D, \mu,\sigma$ should be in principle determined from fits. In
practice, for the purpose of the study some parameters were fixed to
$C=4$, $\mu=0$ GeV, $\sigma=1$ GeV~\cite{Jung:2001hx}. 
The value of parameter $C$ is dictated by the spectator counting
rules\cite{Brodsky:1973kr}. Fit of $\chi^2/\text{n.d.f.}=1.4$ was achieved.

Authors of~\cite{Bacchetta:2010hh} used a grid in the parameter space to parameterise the $ep$ cross section obtained from the {\sc Cascade} Monte Carlo generator and to fit the $ep$ cross section obtained from HERA data. Details of the technique can be found in the publication. 

\section{Forward jets in Monte Carlo generators {\sc Pythia} and {\sc Cascade}}

The analytical results for forward jet production which are implemented in Monte Carlo generator {\sc Cascade} can be found in~\cite{Deak:2009xt}. More complete discussion of numerical results is available in~\cite{Deak:2010gk}.

\subsection{Transverse momentum spectra}
\label{sec:trv}
\begin{figure}[t!]
  \begin{picture}(30,0)
    \put(-15, -190){
      \includegraphics{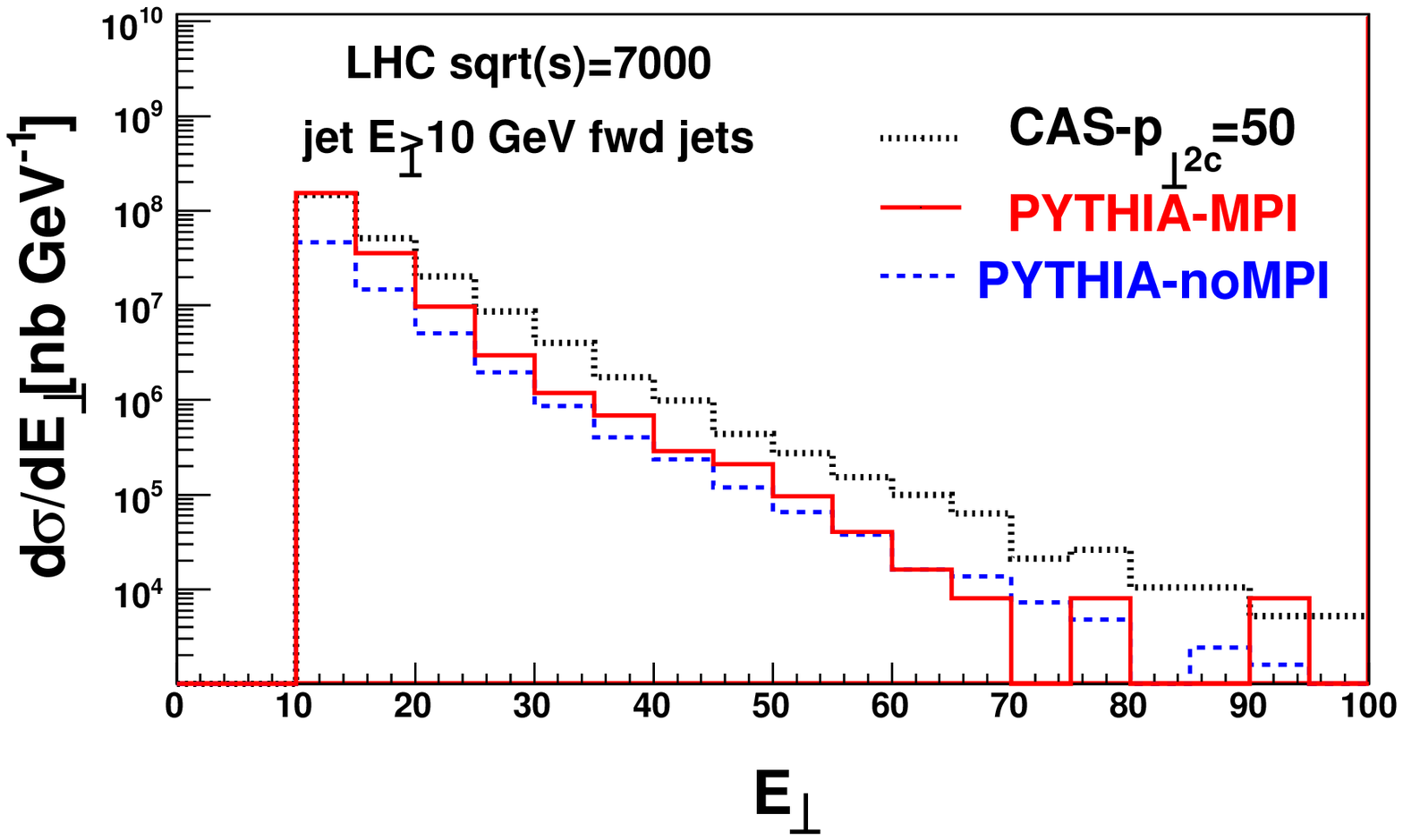}
    }
    \put(200, -190){
      \includegraphics{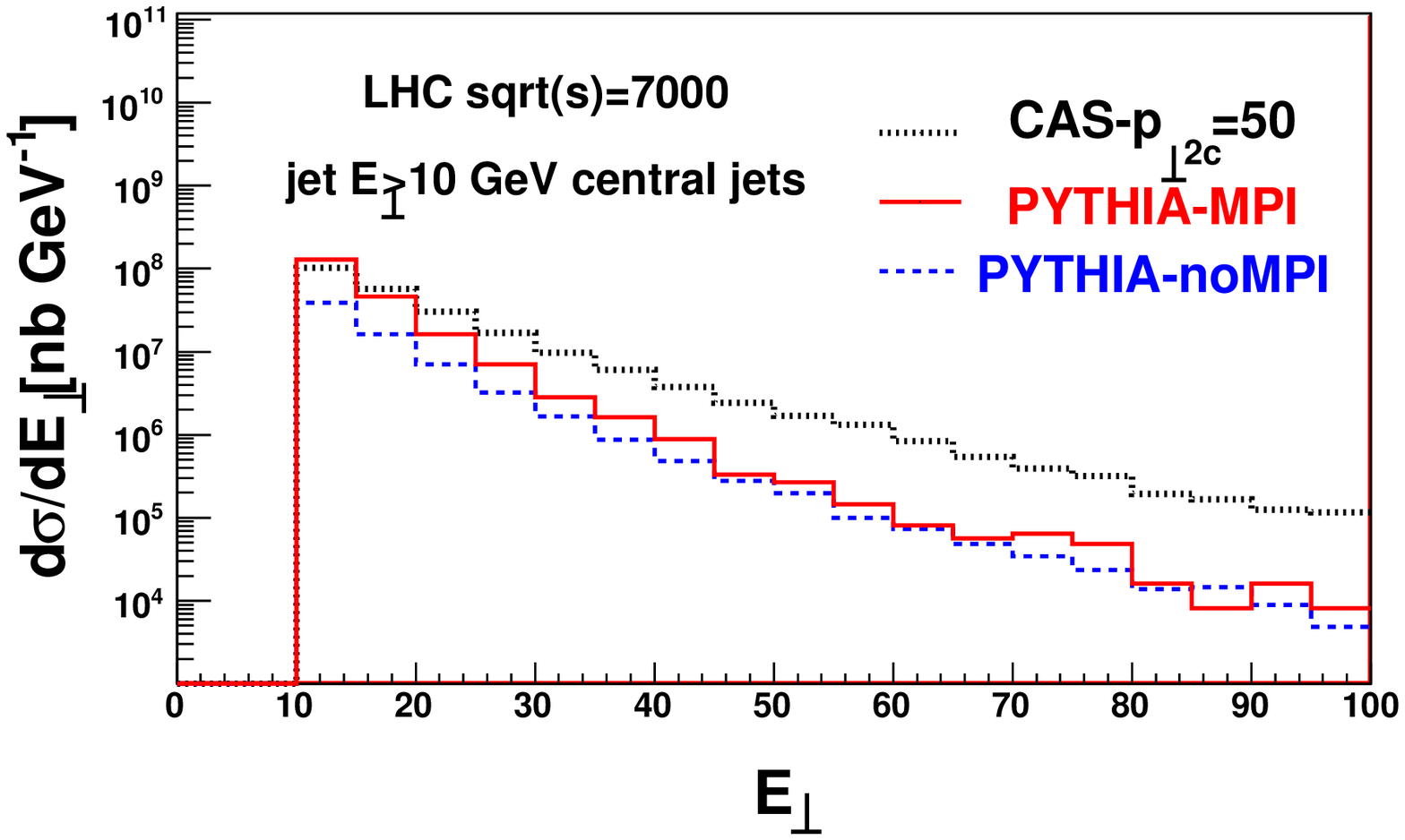}
    }

     \end{picture}
\vspace{6.2cm}
\caption{\em \small Transversal momentum spectra of produced jets at total collision energy $\sqrt s=7\,TeV$ with requirement that p$_\perp\!>\!10\,GeV$. 
We compare predictions obtained from  {\sc Cascade} and {\sc Pythia} running in a multiple interactions mode and no multiple interactions mode. Spectrum of forward jets (left);  
spectrum of central jets (right).}
\label{Fig:transversal}
\end{figure}

In Fig.~\ref{Fig:transversal} the prediction of differential cross section $\frac{d\sigma}{d p_{\perp}}$
is shown as obtained from {\sc Cascade} and {\sc Pythia}. The cross sections predicted from both simulations at low momentum are of the similar order,  
however, at larger transverse momentum the  {\sc Cascade} predicts a 
larger cross section what is clearly visible for central jets (Fig.~\ref{Fig:transversal} right).
This behavior is expected since  {\sc Cascade} uses matrix elements which are calculated within high energy factorization scheme allowing for harder transversal
momentum dependence as compared to collinear factorization. Moreover {\sc Cascade} 
applies CCFM parton shower utilizing angle dependent evolution kernel which at small $x$ does not lead to ordering in transverse momentum, 
and thus allow for more hard radiations during evolution as compared to based on leading order DGLAP splitting functions Monte Carlo generator {\sc Pythia}. 
The parton shower has major influence on the side where the small $x$ gluon enters the hard interaction, thus the jets in the central region are 
mainly affected by the parton shower. 

\subsection{Rapidity dependence}
\label{sec:rap}
\begin{figure}[t!]
  \begin{picture}(30,0)
    \put(-15, -190){
      \includegraphics{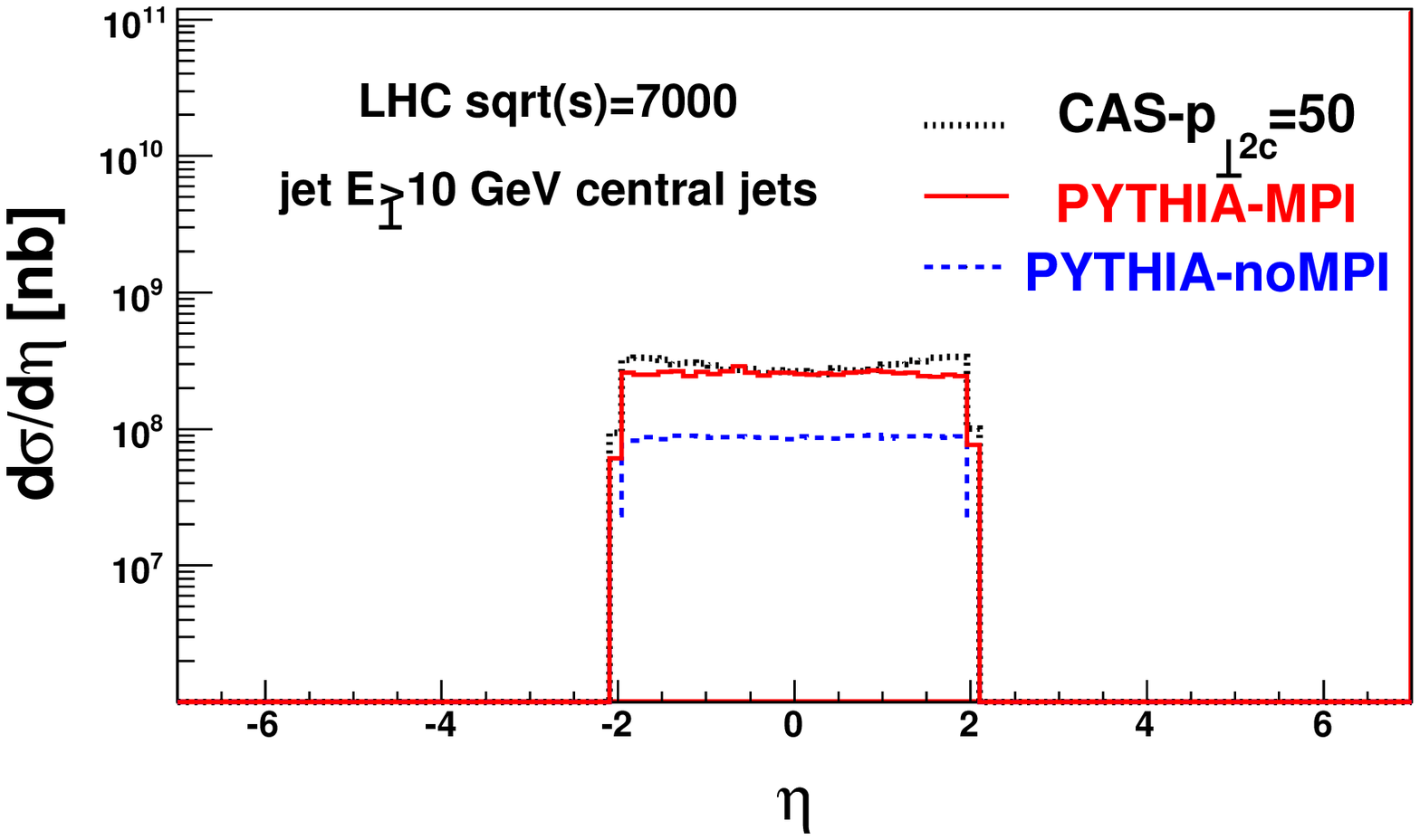}
    }
    \put(200, -190){
      \includegraphics{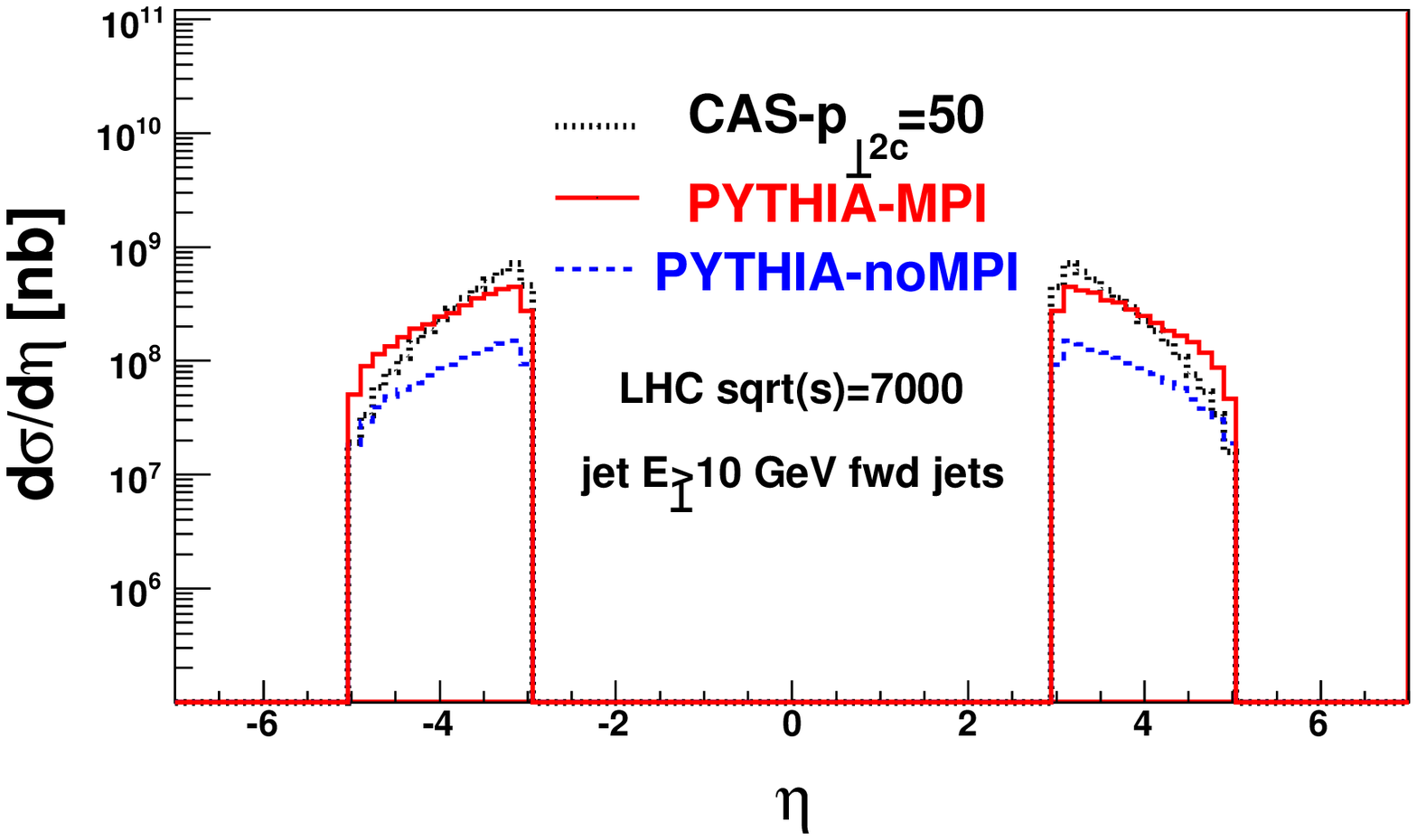}
    } 
     
     \end{picture}
\vspace{6.2cm}
\caption{\em \small  Pseudorapidity spectra of produced jets at total collision energy $\sqrt s=7\,TeV$with requirement that $p_T\!>\!10GeV$. 
We compare predictions obtained from {\sc Cascade} and {\sc Pythia} running in multiple interactions mode and no multiple interactions mode. Spectrum of forward jets
(left); spectrum of central jets (right).}
%\vspace{7cm}
\label{Fig:rapidity}
\end{figure}

In fig.~\ref{Fig:rapidity} we show prediction for pseudorapidity dependence of the cross section in two regions $0\!<\!|\eta|\!<\!2$ and $3\!<\!|\eta|\!<\!5$.
We see that results from {\sc Cascade} interpolate between {\sc Pythia} with multiple interactions in the central region and {\sc Pythia} without multiple interactions 
in the forward region.  
The result is due to the fact that  {\sc Cascade} (because of angular ordering), and {\sc Pythia} with multiple interactions (because of multi chain exchanges), predict 
more hadronic activity  in the central rapidity region as compared to the collinear shower. 
In the remaining rapidity region cascade uses collinear parton shower of a similar type as in {\sc Pythia} without multiple interactions.

\section{$ZQ\bar{Q}$ production in {\sc MCFM} and {\sc Cascade} Monte Carlo generators}

\begin{figure}[t!]
  \begin{picture}(30,0)
    \put(-15, -190){
      \includegraphics{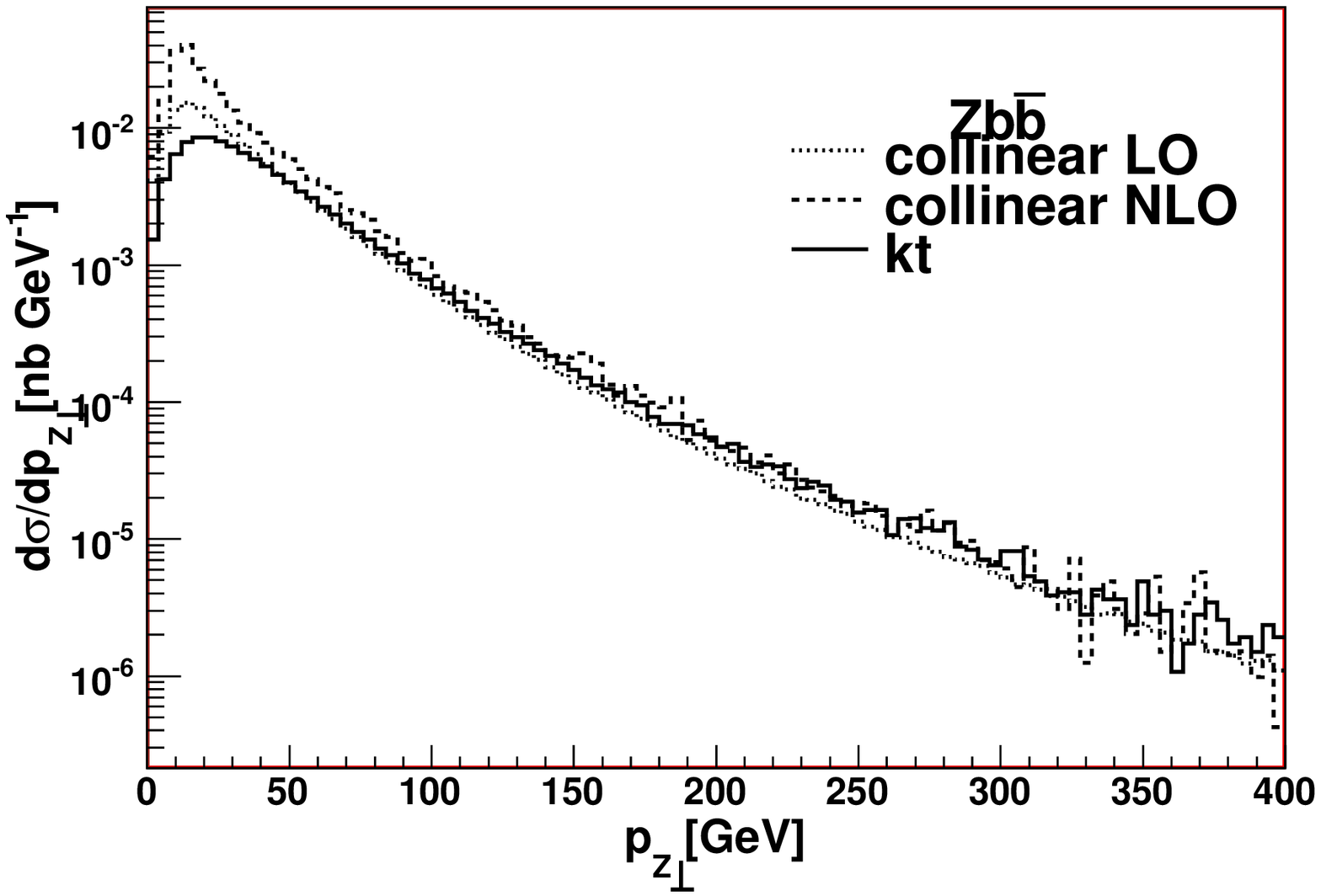}
    }
    \put(200, -190){
      \includegraphics{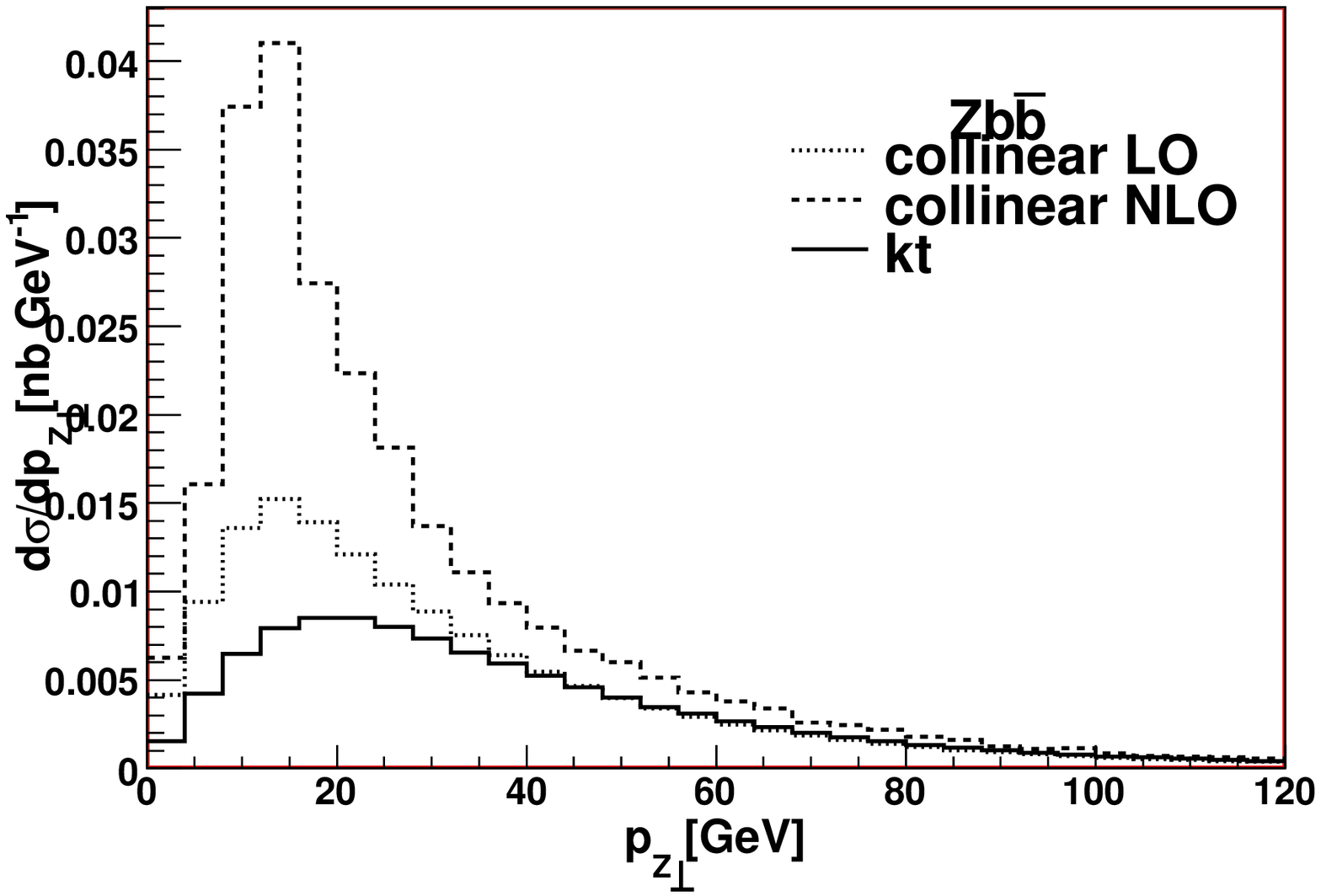}
    } 
     
     \end{picture}
\vspace{6.2cm}
\caption{\em \small Transversal momentum spectrum of the $Z$ boson produced associated with a $b\bar{b}$ pair. Plotted in logarithmic scale (left) and in linear scale (right).}
%\vspace{7cm}
\label{fig:rptZ}
\end{figure}

To compare with a collinear NLO calculation,
we use again the Monte Carlo generator {\sc Mcfm}. This Monte
Carlo generator provides the process $gg \to Zb\bar{b}$ at NLO only in
the massless quark limit. To avoid divergences, additional cuts
are applied on transversal momenta of quarks, on the invariant
mass of the $b\bar{b}$ pair, and on transversal momenta of a gluon
which is produced in diagrams of real NLO corrections. Transversal
momenta of produced quark, antiquark and gluon have to satisfy the
condition $p_\perp>4.62{\rm GeV}$ (corresponding to the mass of the $b$-quark). These
cuts on quark (antiquark) momenta are automatically applied in
{\sc Mcfm} when one is performing a calculation involving massless
quarks (antiquarks). We choose the parton density functions set
CTEQ6M \cite{CTEQ}.
The same cuts on transversal momenta of quark and
antiquark are then applied in {\sc Cascade} as well.

The result for the cross sections differential in the transversal momentum
of $Z$ can be seen in Fig.~\ref{fig:rptZ} left. The cross section
changes especially at small $p_{Z\perp}$ (see
Fig.~\ref{fig:rptZ} right) from LO to NLO calculation, and the
difference between collinear calculation and $k_T$-factorization
calculation becomes more pronounced. We observe that the maximum of the
distribution in the NLO calculation ({\sc Mcfm}) stays approximately
at same value of transversal momenta and the shape of the peak is
very different from the one we obtain in $k_T$-factorization.
Nevertheless, the $p_{Z\perp}$ distributions match at very high
$p_{Z\perp}$ ($\mathcal{O}(10^2 {\rm GeV})$).

The broadening of the peak  by inclusion of small x effects which can be seen in Fig.~\ref{fig:rptZ} is consistent with prediction of~\cite{Berge:2004nt}.

\section{Acknowledgments}

I want to thank Alessandro Bachetta, Hannes Jung, Francesco Hautmann, Albert Knutsson, Krzysztof Kutak and Federico von Samson-Himmelstjerna for having the pleasure to include results of their work in my presentation at XL ISMD 2010 in Antwerp, Belgium.

%\section{Bibliography}

\bibliographystyle{deak_michal}
\bibliography{deak_michal}

\end{document}